\documentclass{aa}
\usepackage{graphicx}
\begin{document}

\title{Diffraction-limited speckle interferometry and modeling of the 
circumstellar envelope of R~CrB at maximum and minimum light}

\author{K.~Ohnaka\inst{1}, Y.~Balega\inst{2}, T.~Bl\"ocker\inst{1}, 
Y.~S.~Efimov\inst{3}, 
K.-H.~Hofmann\inst{1},  N.~R.~Ikhsanov\inst{1}, V.~I.~Shenavrin\inst{4}, 
G.~Weigelt\inst{1}, B.~F.~Yudin\inst{4}
} 

\offprints{K.~Ohnaka, \\ \email{kohnaka@mpifr-bonn.mpg.de}}

\institute{
Max-Planck-Institut f\"{u}r Radioastronomie, 
Auf dem H\"{u}gel 69, D-53121 Bonn, Germany
\and
Special Astrophysical Observatory, Nizhnij Arkhyz, Zelenchuk region, 
35147 Karachai-Cherkesia, Russia
\and
Crimean Astrophysical Observatory, Nauchny, 98409, Crimea, Ukraine, 
and Isaak Newton Institute of Chile, Crimean Branch
\and
Sternberg Astronomical Institute, Universitetskii pr. 13, 119899 Moscow, 
Russia}

\date{Received / Accepted }

\abstract{
We present the first speckle interferometric observations of 
\object{R CrB}, 
the prototype of a class of peculiar stars which undergo 
irregular declines in their visible light curves. 
The observations were carried out with the 6~m telescope at the 
Special Astrophysical Observatory 
near maximum light ($V=7$, 1996 Oct. 1) and at minimum light 
($V=10.61$, 1999 Sep. 28).  
A spatial resolution of 75~mas was achieved in the $K$-band. 
The dust shell around 
R~CrB is partially resolved, and the visibility is approximately 
0.8 at a spatial frequency of 10 cycles/arcsec. 
The two-dimensional power spectra obtained at both epochs 
do not show any significant deviation from circular symmetry. 
The visibility function and spectral energy distribution obtained 
near maximum light can be simultaneously fitted with a model 
consisting of the central star and an optically thin dust shell with 
density proportional to $r^{-2}$.  The inner boundary of the shell is 
found to be $82$~\mbox{$R_{\star}$}\ (19~mas) with a temperature of $920$~K.  
However, this simple model fails to simultaneously reproduce 
the visibility and spectral energy distribution obtained at minimum light.  
We show that this discrepancy can be attributed to thermal 
emission from a newly formed dust cloud. 
}   

\titlerunning{Diffraction-limited speckle interferometry of R CrB}
\authorrunning{K.~Ohnaka et al.}
\maketitle

\section{Introduction}

The R Coronae Borealis (RCB) stars are a class of unusual objects 
characterized by 
sudden declines in their visible light curves as deep as $\Delta V \sim 8$.  
They are extremely hydrogen-deficient and also carbon-rich 
(e.g. Asplund et al. \cite{asplund00} and references therein).  
The RCB stars are thought to undergo the formation of dust clouds 
in random directions, 
and it is believed that a sudden decline takes place, only when a dust 
cloud forms in the line of sight (Loreta \cite{loreta34}, 
O'Keefe \cite{okeefe39}). 
A newly formed dust cloud 
is expected to be accelerated by radiation pressure and to move away, 
expanding and 
dispersing over months, as the object gradually returns to its maximum visual 
brightness. The mechanism of the dust cloud formation and its temporal
evolution are, however, still poorly understood.  
Especially, the location of dust formation is in dispute: far from the star, 
$\ga 20 \mbox{$R_{\star}$}$ (e.g. Fadeyev \cite{fadeyev86}, 
\cite{fadeyev88}, Feast \cite{feast96}), or very close to the 
photosphere, $\sim 2 \mbox{$R_{\star}$}$ (Payne-Gaposchkin \cite{PG63}). 
Recent photometric and spectroscopic observations as well as theoretical 
progress on dust formation suggest that the latter scenario may be 
the case (see, e.g. Clayton \cite{clayton96}, Feast \cite{feast97ii}), 
however, no definitive answer is yet available. 

The spectral energy distributions (SEDs) of RCB stars exhibit 
infrared emission 
peaks around 6 $\sim$ 8~\mbox{$\mu$m}. The IR excess, which accounts for 
typically 30\% of the total flux, is constantly present, 
regardless of the visual brightness of the central star. 
Therefore, the IR excess originates not from a single newly formed 
dust cloud, but mainly from a group of dispersed dust clouds 
with temperatures of approximately 600 -- 900~K.  
For example, 
Walker et al. (\cite{walker96}) fit the infrared (spectro)photometric 
data of R CrB with a 650~K blackbody.  
The study of IRAS observations by Gillett et al. (\cite{gillett86}) 
led to the detection of an additional, very extended ``fossil'' shell 
around R CrB, whose diameter is as large as 18\arcmin, 
with a temperature of $\sim 30$~K.  
Walker (\cite{walker94}) found such fossil shells for at least 
four RCB stars.

Apart from the detection of the fossil shells, most of the observational 
results on the 
circumstellar environment around RCB stars were obtained by photometry 
and spectroscopy.  Recently, Clayton \& Ayres (\cite{clayton01}) have 
revealed extended \ion{C}{ii} $\lambda$1335 emission around two RCB stars, 
\object{V854~Cen} and \object{RY~Sgr}, by long-slit spectroscopy.  
However, such direct information on 
the spatial distribution of material in the vicinity of the central star 
has been very rare up to now. 
In this paper, we present high-resolution speckle interferometry carried out 
for R~CrB at maximum and minimum light.  
The properties of the warm dust shell will be derived by 
simultaneous fits of the observed visibilities and SEDs using 
power-law models.  
We will also discuss the possible indication of a newly formed 
hot dust cloud.

\section{Speckle interferometric observations}
\label{sec_speckleobs}

\begin{figure}
\resizebox{\hsize}{!}{\rotatebox{-90}{\includegraphics{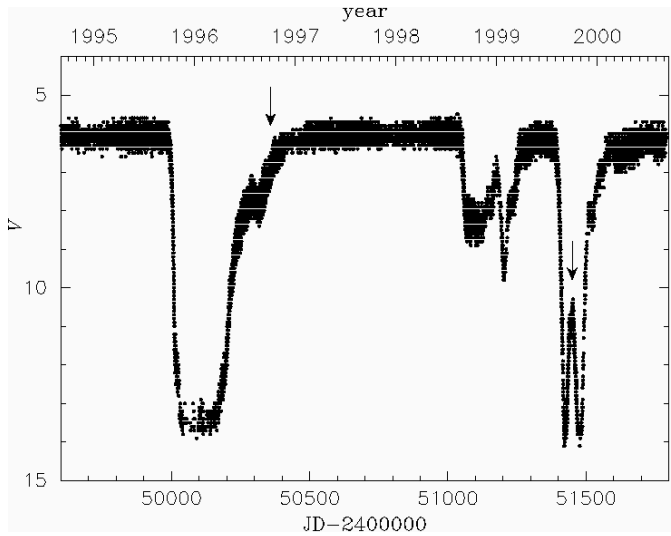}}}
\caption{Visual light curve of R CrB based on the AAVSO data.  
The epochs of the speckle interferometric observations are shown by 
the arrows}
\label{lightcurve}
\end{figure}

\begin{table}
\caption[]{Speckle interferometric observations.  
$\lambda_{c}/\Delta \lambda$: central wavelength and 
FWHM bandwidth of the filters, 
$N_{\rm T}$: number of speckle interferograms acquired for the target,   
$N_{\rm R}$: number of speckle interferograms acquired for the 
reference stars, 
$T$: exposure time of each frame, 
$S$: seeing, and 
$p$: pixel size
}
\label{obslog}
\begin{flushleft}
\begin{tabular}{ccc}
\hline
\noalign{\smallskip}
           &  1996 Oct. 1   &  1999 Sep. 28  \\
\noalign{\smallskip}
\hline   
\noalign{\smallskip}
JD         &  2450358.2     & 2451450.2  \\
$V$ (mag)  &  7           & 10.61   \\
$\lambda_{c}/\Delta \lambda$(\mbox{$\mu$m})  &  2.191/0.411 & 2.115/0.214 \\
Reference star  & \object{HIP80322}  & \object{HIP77743} \\
 $N_{\rm T}$  &  167 &  990 \\
 $N_{\rm R}$  &  198 & 1044 \\
 $T$(ms)      &  150 &  80 \\
 $S$(arcsec)  &  1.4 &  1.1 \\
 $p$(mas)     &  30.5 & 26.4 \\
Field of view & 8\farcs9$\times$8\farcs9 & 5\farcs1$\times$5\farcs1 \\
\noalign{\smallskip}
\hline
\end{tabular}

\end{flushleft}
\end{table}

\begin{figure}
\sidecaption
\resizebox{\hsize}{!}{\includegraphics[width=12cm]{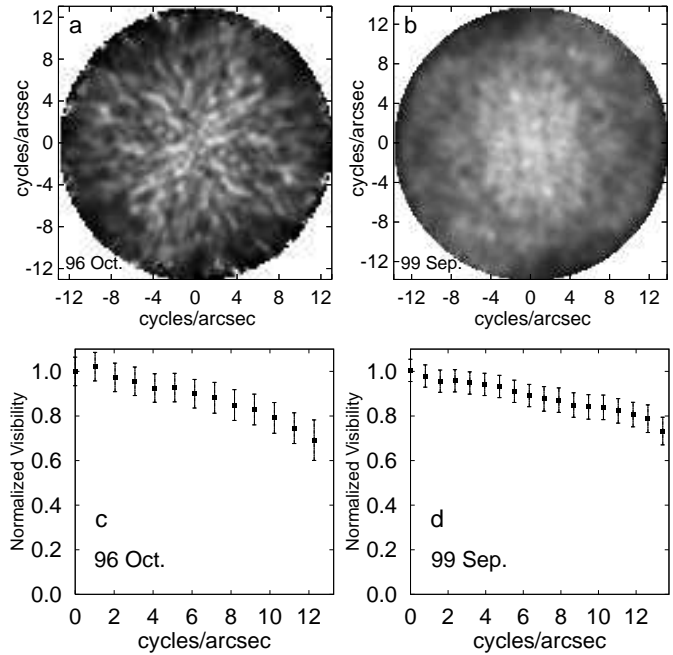}}
\caption{ {\bf a} and {\bf b} Two-dimensional power spectra of R CrB 
obtained on 1996 October 1 and on 1999 September 28, respectively.  
{\bf c} and {\bf d} 
Azimuthally averaged visibility functions derived from {\bf a} and {\bf b}, 
respectively}
\label{obsspeckle}
\end{figure}

The $K$-band speckle interferometric 
observations were carried out with the 6~m telescope at the Special 
Astrophysical Observatory (SAO) in Russia, using our NICMOS-3 camera for the 
observation in 1996 and our HAWAII array speckle camera in 1999. 
The observations are described in Table~\ref{obslog}.  

Fig.~\ref{lightcurve} shows the visual light curve of R CrB in the
relevant period based on the compiled data of the American Association 
of Variable Star Observers (AAVSO).  The data before 1996 were 
taken from AAVSO Monograph 4 Supplement 1, while J.~A.~Mattei kindly 
provided unpublished data for the rest of the period.  
On 1996 October 1, R CrB was on its final
recovery from the 1995-96 deep minimum and its visual magnitude was 
around 7 (Mattei \cite{mattei00}).  Since $V$ is approximately 6 
at maximum light, the star was slightly obscured by $\Delta V = 1$.  
On 1999 September 28, 
the star was just between two sharp deep minima and had a visual 
magnitude of 10.61 (see Sect.~\ref{sec_photo}).  The star was 
heavily obscured by $\Delta V = 4.6$.

Fig.~\ref{obsspeckle} shows the two-dimensional power spectra and
azimuthally averaged visibility functions reconstructed with the speckle 
interferometry method (Labeyrie \cite{labeyrie70}).  
The error bars include systematic errors caused by seeing 
variations as well as speckle noise errors.  
There is no significant deviation from circular symmetry 
in the two-dimensional power spectra observed at both epochs.  
The granular features seen in Fig.~\ref{obsspeckle}a are the speckle 
noise resulting from the small number of interferograms.  
In Fig.~\ref{obsspeckle}b, a slight elongation along the vertical axis 
of the figure can be marginally recognized.  
However, this is within the error bars, 
and cannot be regarded as a significant feature.  
Clayton et al. (\cite{clayton97}) suggest a bipolar geometry for R CrB 
based on the wavelength dependence of the position angle of polarization. 
Our speckle observations seem to support a rather symmetric distribution 
of material.  However, a disc or torus may lie very 
close to the star, and therefore, it is premature to rule out a bipolar 
geometry completely.  

The reconstructed visibility functions shown in Figs.~\ref{obsspeckle}c 
and ~\ref{obsspeckle}d exhibit no difference larger than the error bars. 
The source probably consists of the central star (a point source) 
and an extended dust shell.  
Visibilities observed for such objects 
have a plateau at high spatial frequencies resulting from the point 
source.  In the observations presented here, the angular size of 
the shell is small, and therefore, the plateau is not visible 
within a cut-off frequency of $\sim$ 13~cycles/arcsec. 
The determination of the physical parameters of the shell by 
simultaneous fits to the observed 
visibilities and SEDs will be presented in Sect.~\ref{simul_fit}.

\section{Photometric data}
\label{sec_photo}

$UBVRIJHKLM$ photometry was carried out 
on 1999 September 29, just one night after the speckle observation.  
The optical ($UBVRI$) data were obtained with the 1.25~m telescope at 
the Crimean Astrophysical Observatory, and the infrared ($JHKLM$) data 
with the 1.22 m telescope at the Crimean Laboratory of the Sternberg 
Astronomical Institute.  

No photometric data, except for the visual 
magnitudes compiled in the AAVSO database, 
have been published around the date of our speckle observation in 
1996 October. 
Therefore, we use the data published by Shenavrin et al. 
(\cite{shenavrin79}), who report a series of photometric observations 
during the 1977 deep minimum.  We use the $UBVRJHK$ data on 1978 
January 22 (JD2443530.6), when the visual 
magnitude was 7.03, almost the same as that on the 
date of our speckle observation in 1996 October.  
For the $L$- and $M$-bands, we use photometric data obtained on 1996 
June 11 (JD2450246.4) by Shenavrin (unpublished observation), 
about 4 months before our speckle observation.  
It is known that the variation of the $L$- and $M$-band fluxes has no
correlation with the visual light curve, and that they exhibit 
a semi-periodic variation of $\sim 1.5$~mag with 
a period of about 1260~days (e.g. Feast et al. \cite{feast97}).  
In fact, the $L$ and $M$ magnitudes on 1996 June 11 are by 1.09~mag and 
1.25~mag brighter than those on 1999 September 29, respectively. 
The $L$- and $M$-band fluxes measured 4 months before are expected to 
represent the values on the date of the speckle observation rather well.  

We also use IRAS observations at 12, 25, 60, and 100~\mbox{$\mu$m}.  
The very extended ``fossil'' shell 
gives rise to flux excess at 60 and 100 \mbox{$\mu$m}, when evaluated 
with a large aperture (Gillett et al. \cite{gillett86}).  
In this paper, however, we are primarily interested in 
the warm dust shell which was not resolved with IRAS, therefore, 
we adopt the point source processed fluxes for 60 and 100 \mbox{$\mu$m}.  
Long-term variations at wavelengths longer than 10~\mbox{$\mu$m}\ are 
considered to be smaller than those in the $L$- and $M$-bands.  
Forrest et al. (\cite{forrest72}) show that the amplitude at 11~\mbox{$\mu$m}\ 
is about half as small as those in the $L$- and $M$- bands.  
The variation at wavelengths longer than 11~\mbox{$\mu$m}\ is estimated 
to be $\sim 0.75$~mag at most.  

The interstellar extinction toward R CrB is $E(B-V) = 0.05$ 
(Asplund et al. \cite{asplund97}).  The observed fluxes are dereddened 
using the method of Savage \& Mathis (\cite{savage79}) with 
$A_{V} = 3.1 E(B - V)$.

\section{Description of the model}
\label{sec_model}

\begin{figure}
\resizebox{\hsize}{!}{\rotatebox{-90}{\includegraphics{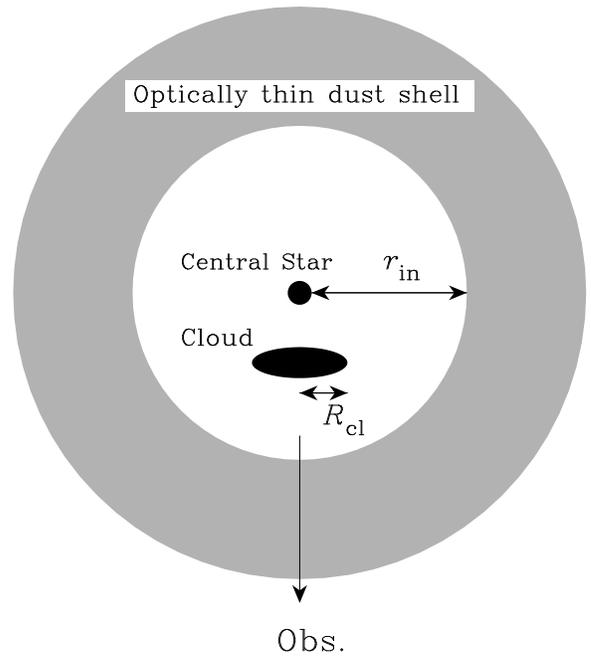}}}
\caption{Schematic view of the central star, 
an obscuring dust cloud, and an optically thin dust shell.  
The figure is not to scale}
\label{inset}
\end{figure}

The basics of the model used here are the same as adopted 
by Gillett et al. (\cite{gillett86}).  
Fig.~\ref{inset} illustrates the picture considered in our modeling.  
It consists of the central star and an optically thin dust shell
responsible for the constant IR excess.  
As mentioned in Sect.~\ref{sec_speckleobs}, the star was obscured 
by $\Delta V = 4.6$ (1999 September 28) and $\Delta V = 1.0$ (1996
October 1).  Therefore, an additional obscuring dust cloud, 
which is assumed to be circular 
(projected onto the sky as seen from the star), 
is also considered, as shown in Fig.~\ref{inset}.  

Adopting $M_{\rm bol}$ and \mbox{$T_{\rm eff}$}\ to be $-5.3$ 
(Gillett et al. \cite{gillett86}) and 6600 -- 6900~K 
(Asplund et al. \cite{asplund00}), respectively, 
the radius of the central star is approximately 
70~\mbox{$R_{\sun}$}.  We adopt 6750~K for the effective temperature, and 
radiation from the central star is approximately 
represented by a blackbody.  
The distance to R CrB 
adopted here is 1.6~kpc (Gillett et al. \cite{gillett86}), 
yielding an angular radius of the central star of 0.23~mas.  

We consider a spherical shell with an inner radius \mbox{$r_{\rm in}$}\ and outer 
radius \mbox{$r_{\rm out}$}.  
The shell is assumed to be optically thin not only in the infrared but
also in the optical, since there 
is no evidence for strong circumstellar reddening for R CrB.  Asplund et al. 
(\cite{asplund97}) found that the observed optical flux at maximum light 
can be well reproduced by their line-blanketed model atmospheres without 
any circumstellar extinction.  In fact, the optical depth of the dust shell 
in the $V$-band is $\la 0.3$ in our models presented below.  
The assumption of spherical symmetry for the shell is 
justified, given the 
negative detection of deviation from circular symmetry in the 
observed two-dimensional power spectra.  

The grain number density is 
assumed to decrease in a power-law of radius, 
\begin{equation}
n (r) = C r^{- \gamma},   
\label{denseq}
\end{equation}
where $C$ is a constant which does not affect the shape of emergent 
SEDs, and therefore, 
should be adjusted to fit the flux observed on the earth. 

\begin{figure}
\resizebox{\hsize}{!}{\rotatebox{-90}{\includegraphics{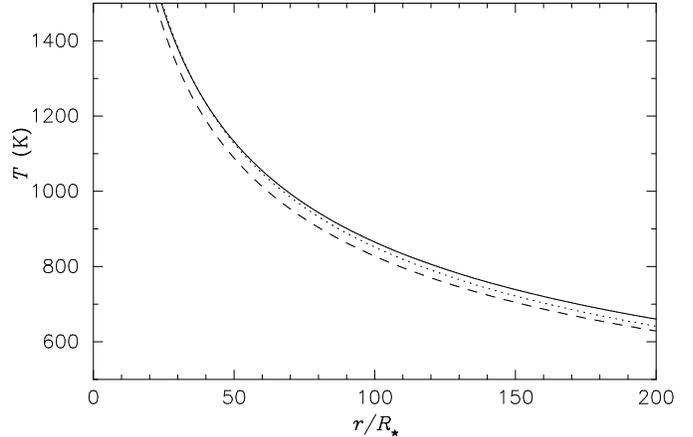}}}
\caption{Temperature distributions derived from the thermal balance equation. 
Dotted line: Colangeli et al. (\cite{colangeli95}), dashed line: 
Bussoletti et al. (\cite{bussoletti87}), solid line: Rouleau \& Martin 
(\cite{rouleau91}) with $a = 0.01$~\mbox{$\mu$m}}
\label{Tradeq}
\end{figure}

The temperature 
in an optically thin dust shell is determined by the thermal balance 
equation 
\begin{eqnarray}
\int_{0}^{\infty} \frac{L_{\nu}}{4 \pi r^2} \pi a^2 Q_{\rm abs}(a,\nu) 
d\nu  \nonumber\\
= \int_{0}^{\infty} 4 \pi a^2 Q_{\rm abs}(a,\nu) \pi B_{\nu}(T(r))d\nu , 
\label{energyeq}
\end{eqnarray}
where $L_{\nu}$ is the luminosity of the central star at a given
frequency, $a$ is the radius of a grain, $Q_{\rm abs}(a,\nu)$ 
is the absorption efficiency of a grain, and $B_{\nu}(T(r))$ is 
the Planck function.  
Amorphous carbon is the most probable candidate for the circumstellar
dust around RCB stars (e.g. Holm et al. \cite{holm82}, 
Hecht et al. \cite{hecht84}).  
In Fig.~\ref{Tradeq}, we show temperature distributions predicted 
by the thermal balance equation, using the extinction of amorphous
carbon obtained by Bussoletti et al. (\cite{bussoletti87}) (AC2 sample), 
Rouleau \& Martin (\cite{rouleau91}) (AC1 sample), and Colangeli et al. 
(\cite{colangeli95}) (ACAR sample).   
We calculate $Q_{\rm abs}$ from the complex 
refractive index derived by Rouleau \& Martin (\cite{rouleau91}) 
in the Mie theory, assuming a single grain size, $a = 0.01$~\mbox{$\mu$m}, 
and using the code published by Bohren \& Huffman (\cite{bohren83}).  
This grain size is based on the result obtained by 
Hecht et al. (\cite{hecht84}), who analyzed the UV spectra of R~CrB 
and RY~Sgr and concluded that the grain size is between 
0.005 and 0.06~\mbox{$\mu$m}.  
As Fig.~\ref{Tradeq} shows, the temperature distributions agree with one 
another within $\sim 50$~K.  

The flux density observed on the earth from the optically thin shell 
described above can be calculated by 
\begin{equation}
f_{\rm s} (\lambda) = \frac{4 \pi m_{\rm d} \kappa_{\lambda} C}{D^2} 
\int_{r_{\rm in}}^{r_{\rm out}} \!\!\!  B_{\lambda} (T(r)) r^{2-\gamma} \,dr ,
\label{fluxeq}
\end{equation}
where $m_{\rm d}$ and $\kappa_{\lambda}$ is the mass and mass absorption 
coefficient of a grain, respectively, and $D$ is the distance to the star.  
The constant $C$ is adjusted so that the flux predicted from the 
models can reproduce the observed flux at 25~\mbox{$\mu$m}, since 
its long-term variation mentioned in 
Sect.~\ref{sec_photo} is expected to be minimal, 
and at the same time, 
it does not exhibit the contribution of the fossil shell.  
The variation of the 25~\mbox{$\mu$m}\ flux is not available in the literature, 
but it is estimated to be smaller than the 0.75~mag at 11~\mbox{$\mu$m}.  
$r_{\rm out}$ is set to be $\sim 1.5 \times 10^4$~\mbox{$R_{\star}$}, which gives a good 
match to the far-infrared part of the observed SEDs, as we will show
below.

It should be stressed here that the real circumstellar environment 
around R CrB is most likely much more complex than depicted by 
the above spherical shell model.  
Since dust ejection in RCB stars presumably occurs in clouds, 
not in a spherical shell, it is likely that the real distribution of 
material is clumpy or patchy without any clear inner boundary.  
At large distances, 
however, it is still plausible that a spherical shell model roughly 
represents the distribution of material, 
as long as dust clouds are ejected randomly and frequently.  
There may be some newly formed clouds inside the inner boundary 
defined in the above spherical model, if the star ejects dust clouds 
frequently, for example, every pulsational cycle 
(40 -- 50 days).  Without detailed knowledge about the 
dispersal processes of clouds, however, it is beyond the scope of this paper 
to construct a more detailed model.  In the framework of our models, 
we consider only one newly formed dust cloud, as Fig.~\ref{inset} 
illustrates.  

\begin{figure}
\resizebox{\hsize}{!}{\rotatebox{0}{\includegraphics{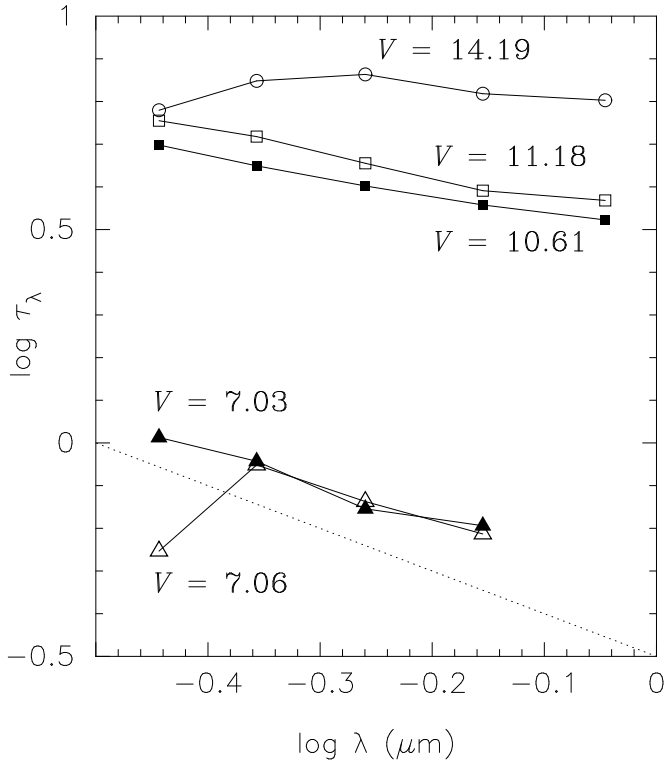}} }
\caption{Extinction curves of a dust cloud and its temporal variation.  
The open symbols represent the extinction curves derived from  
the observations covering the 1983 minimum obtained by 
Goncharova (\cite{goncharova92}). 
Open circles: 1983 Oct. 17 (JD2445625.21), open squares: 
1984 Jan. 16 (JD2445715.60), and open triangles: 1984 May 22 
(JD2445843.43).  The filled squares represent the extinction 
derived for 1999 Sep. 29.  The filled triangles 
represent the extinction derived from the data obtained on 1978 Jan. 22 
by Shenavrin et al. (\cite{shenavrin79}).  
The visual magnitude at each epoch is also given.  
The extinction characterized by $1/\lambda$ is plotted with 
the dotted line for reference}
\label{extfig}
\end{figure}

The observed SEDs (see Sect.~\ref{simul_fit}) 
demonstrate that the contribution of the central star is not negligible 
in the near-infrared.  
The flux from the central star was attenuated by a dust cloud 
in front of the star by $\Delta V \sim 1$ on 1996 October 1 
and $\Delta V \sim 4.6 $ on 1999 September 28.  
In order to estimate the attenuated flux from the star in the near-infrared, 
the extinction curve 
of the dust cloud is empirically derived from photometry in the 
optical, where emission from the dust shell is negligible. 
The effect of the obscuration due to a dust cloud is expressed as 
\begin{equation}
f_{\star}(\lambda) = f_{\star}^{\rm max} (\lambda) 
\exp({-\tau^{\rm cl}(\lambda)}),
\label{obscureeq}
\end{equation}
where $f_{\star}(\lambda)$ and $f_{\star}^{\rm max} (\lambda)$ denote 
fluxes observed at any given time and at maximum light, respectively, 
and $\tau^{\rm cl}(\lambda)$ is the optical depth of the dust cloud 
in front of the star. 
Note that the effect of interstellar extinction, though small for R CrB, 
cancels out, because it affects both $f_{\star}(\lambda)$ and 
$f_{\star}^{\rm max} (\lambda)$ by the same amount.  

Fig.~\ref{extfig} shows the temporal change of the 
optical depth of a dust cloud during the 1983 minimum in the $U$, $B$, $V$, 
$R$, and $I$ bands.  
The optical depth in each band was derived using the photometric 
data in the decline and at maximum light 
obtained by Goncharova (\cite{goncharova92}). 
We also plot the extinction curves derived for the photometric data 
discussed in Sect.~\ref{sec_photo}.  
The figure illustrates that the extinction is almost independent of 
the wavelength at the very bottom of the deep minimum ($\mbox{$V$} = 14.19$).  
As the star starts 
its final recovery, the extinction curve starts to steepen. 
Pugach (\cite{pugach84}) proposed that the neutral extinction 
observed during the initial drop to minimum light can be explained 
by an optically very thick dust cloud whose coverage over the stellar disk 
varies with time.  
On the other hand, Hecht et al. (\cite{hecht84}) analyzed the 
change of the ratio of total to selective extinction ($A_V/E(B-V)$) 
during decline events for R CrB and another RCB star, RY Sgr, 
and suggested that rather large glassy
carbon particles with radii from 0.075 to 0.15~\mbox{$\mu$m}\ might be 
responsible for the nearly neutral extinction seen at the very bottom of 
minimum light.  
These sizes are significantly larger than the 
0.01~\mbox{$\mu$m}\ we adopt for a single grain size, but it should be noted 
that we adopt 0.01~\mbox{$\mu$m}\ as the grain size in the optically thin dust 
shell, while the large grains mentioned above are claimed to be present 
in a newly formed optically thick dust cloud.  
For the temporal change of the extinction at the rise phase from 
minimum light, Hecht et al. (\cite{hecht84}) proposed 
shattering of large grains as a possible mechanism.  
Efimov (\cite{efimov90}) also proposed that the observed temporal 
variation of colors and polarization 
can be accounted for by the change of grain sizes, 
but assuming graphite grains instead of amorphous carbon.  
In any case, the spectral index of the dust cloud approaches 1, as the 
cloud disperses and becomes part of the optically thin dust shell.  
Therefore, the dust properties in the optically thin shell may presumably 
remain constant.

The lack of detailed knowledge about the properties of dust grains 
formed in R CrB and of their temporal variation 
forces us to adopt an empirical extinction law.  
As Fig.~\ref{extfig} 
shows, the extinction due to a dust cloud can be approximated by 
$\tau^{\rm cl}(\lambda) \propto  (1/\lambda)^p $. 
At the bottom of minimum light, $p$ is approximately 0, 
and changes to $\sim 1$, as the star returns to maximum light.  
By the least square fit for the extinction curves shown 
in Fig.~\ref{extfig}, we derive $\log \tau^{\rm cl}(\lambda)  = 
0.52 - 0.47 \log \lambda (\mbox{$\mu$m})$ 
when the $V$ magnitude of the star is 10.61, 
while $\log \tau^{\rm cl}(\lambda) = -0.18 - 0.84 \log \lambda (\mbox{$\mu$m})$
when the $V$ magnitude is 7.03.  
The extinction at longer wavelengths is then estimated by extrapolation. 
Obviously this procedure is not optimal, but with no alternatives 
at hand for disentangling the contributions of the central star and 
the dust shell for the data studied here, we are forced to 
adopt this method.  

Assuming a blackbody for the flux of the 
central star, the flux density observed on the earth from the 
obscured central star is 
\begin{equation}
f_{\star} (\lambda) = \left(\frac{\mbox{$R_{\star}$}}{D}\right)^2 \pi B_{\lambda} (\mbox{$T_{\rm eff}$}) 
\times \exp({-\tau^{\rm cl}(\lambda)} - \tau^{\rm s}(\lambda)),
\label{flstar}
\end{equation}
where $\tau^{\rm s}(\lambda)$ is the optical depth of the dust 
shell along the radial direction.

The flux density observed toward 
R CrB is, therefore, the sum of the flux from the central star obscured 
by a dust cloud and slightly dimmed by the optically thin dust shell 
(equation (\ref{flstar})), thermal emission from the 
optically thin dust shell (equation (\ref{fluxeq})), and 
the thermal emission of a newly formed (therefore presumably still 
optically thick) dust cloud given by 
\begin{equation}
f_{\rm cl} (\lambda) = \left(\frac{\mbox{$R_{\rm cl}$}}{D}\right)^2 \pi 
B_{\lambda} (\mbox{$T_{\rm cl}$}) \times \exp(- \tau^{\rm s}(\lambda)),
\label{flcloud}
\end{equation}
where \mbox{$R_{\rm cl}$}\ and \mbox{$T_{\rm cl}$}\ are the radius and the temperature of the 
newly formed dust cloud, respectively.

The intensity distribution for the central star and the dust shell 
can be written as 
\begin{eqnarray}
I_{\lambda} (b) & = & 
B_{\lambda}(T_{\star}) \exp(-\tau^{\rm cl}(\lambda)-\tau^{\rm s}(\lambda)) 
\times {\rm circ}(b/\mbox{$R_{\star}$})\nonumber\\
            & + & 
2 m_{\rm d} \kappa_{\lambda} \int_{z_{\rm min}}^{\sqrt{r_{\rm out}^2-b^2}} 
\!\!\! n(r) B_{\lambda}(T(r)) \, dz\nonumber\\
            & + &
B_{\lambda}(\mbox{$T_{\rm cl}$}) \exp(-\tau^{\rm s}(\lambda)) 
\times {\rm circ}(b/\mbox{$R_{\rm cl}$}),
\end{eqnarray}
where $b$ is the impact parameter and $z=\sqrt{r^2-b^2}$.  The function 
${\rm circ}(b/\mbox{$R_{\star}$})$ takes a value of 1 for $|b| < \mbox{$R_{\star}$}$ and 0 
elsewhere.  
The lower limit of the integration is 
$z_{\rm min} = 0$ for $b \geq \mbox{$r_{\rm in}$}$, while 
$z_{\rm min} = \sqrt{r_{\rm in}^2-b^2}$ for $b < \mbox{$r_{\rm in}$}$. 
The visibility is calculated by taking the modulus of 
the Fourier transform of the intensity distribution.

\section{Simultaneous fit of observed visibilities and SEDs}
\label{simul_fit}

\subsection{Model fitting without thermal emission from a dust cloud}
\label{easyfit}

\begin{figure}
\resizebox{\hsize}{!}{\rotatebox{0}{\includegraphics{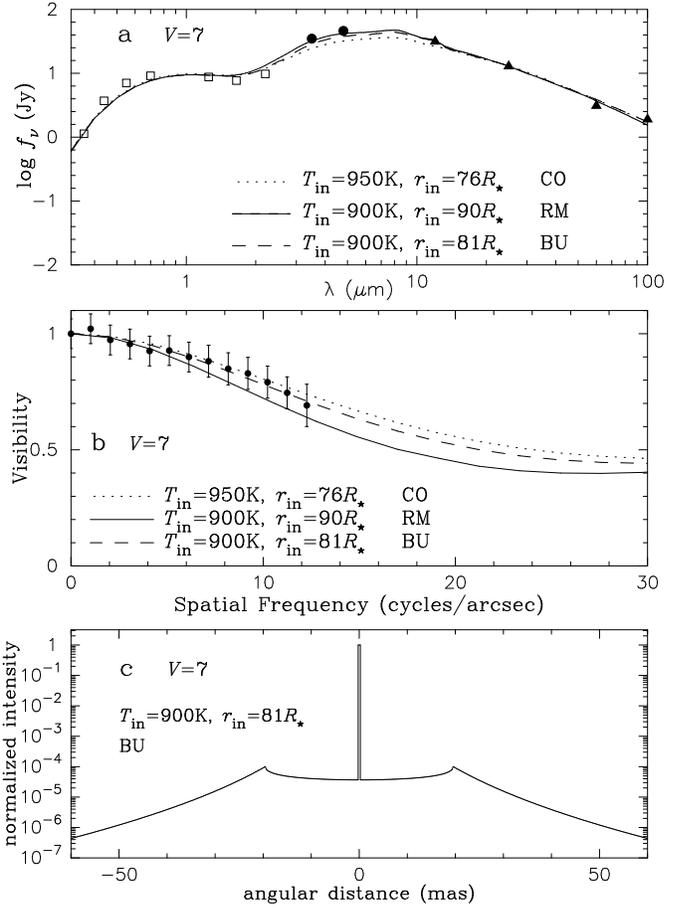}}}
\caption{Simultaneous fit of the SED and visibility of R CrB 
observed on 1996 October 1, using the models discussed 
in Sect.~\ref{easyfit}.  
{\bf a} The open squares, filled circles, and 
filled triangles represent the photometric data obtained 
by Shenavrin et al. (\cite{shenavrin79}), 
Shenavrin (unpublished observation), and 
IRAS, respectively. The three curves represent the SEDs predicted 
from models.  CO represents the model with the data derived by 
Colangeli et al. (\cite{colangeli95}), RM with those derived by 
Rouleau \& Martin (\cite{rouleau91}), and BU with those derived by 
Bussoletti et al. (\cite{bussoletti87}).  
{\bf b} The filled circles represent the observed visibility, 
while the three curves represent the visibilities predicted from the models. 
{\bf c} Normalized intensity profile ($\lambda = 2.2~\mbox{$\mu$m}$) 
of the BU model}
\label{risebestfit}
\end{figure}

\begin{figure}
\resizebox{\hsize}{!}{\rotatebox{0}{\includegraphics{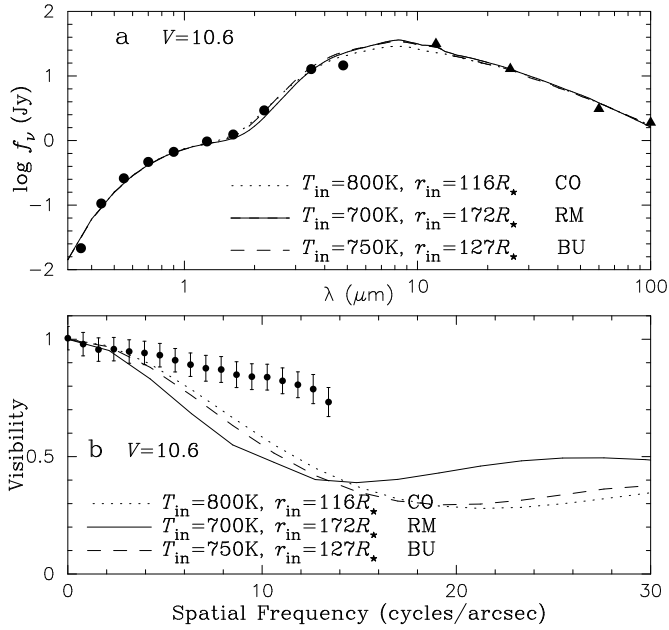}} }
\caption{Simultaneous fit of the SED and visibility of R CrB 
observed on 1999 September 28, using the models without thermal emission 
from the newly formed dust cloud, as discussed in Sect.~\ref{easyfit}. 
{\bf a} The filled circles 
represent the photometric data obtained 
one night after the speckle observation.  
The IRAS data are represented by the filled triangles. 
The three curves represent the SEDs predicted from models.  
See also the legend to Fig.~\ref{risebestfit}.  
{\bf b} The filled circles represent the observed visibility, 
while the three curves represent the visibilities predicted from the models}
\label{midbestfit}
\end{figure}

We first try to fit the observed SEDs and visibilities using 
models {\em without thermal emission from a dust cloud}, namely, 
neglecting the term given by equation (\ref{flcloud}).  
We adopt $\gamma = 2$, 
appropriate for a constant mass loss and expansion velocity.  
The sensitiveness of the near- and mid-infrared parts of emergent SEDs 
to \mbox{$T_{\rm in}$}\ allows us to determine it by fitting the observed SEDs.  
The visibility functions, 
which directly reflect the spatial extent of the shell, 
enable us to examine the validity of the models described above.  

Figs.~\ref{risebestfit} and \ref{midbestfit} 
show the fit of the observed SEDs and visibility functions at 
the two epochs of our speckle observations.  The SEDs and visibilities 
are calculated with the opacities of amorphous carbon obtained by 
Bussoletti et al. (\cite{bussoletti87}) (AC2 sample), 
Rouleau \& Martin (\cite{rouleau91}) (AC1 sample), and Colangeli et al. 
(\cite{colangeli95}) (ACAR sample).   The corresponding temperature 
distributions shown in Fig.~\ref{Tradeq} are used in the calculations.  
Figs.~\ref{risebestfit}a and ~\ref{risebestfit}b demonstrate that 
the SED and visibility for the 
1996 data are well reproduced with \mbox{$r_{\rm in}$}\ = 76 -- 90~\mbox{$R_{\star}$}\ and 
\mbox{$T_{\rm in}$}\ = 900 -- 950~K.  For a given opacity data set, the uncertainty 
of \mbox{$T_{\rm in}$}\ is estimated to be $\pm 100$~K, translating to an uncertainty of 
\mbox{$r_{\rm in}$}\ of $\pm 20$~\mbox{$R_{\star}$}.  Taking the average of the \mbox{$T_{\rm in}$}\ and \mbox{$r_{\rm in}$}\ 
derived with three different opacities and adding the uncertainties 
resulting from the fitting, 
we derive \mbox{$T_{\rm in}$}\ = $920 \pm 103$~K and \mbox{$r_{\rm in}$}\ = $82 \pm 23$~\mbox{$R_{\star}$}.  
Fig.~\ref{risebestfit}c illustrates 
the normalized intensity profile of the best fit model for the 1996 data.  
It consists of the central star and a ring-like structure characteristic 
of an optically thin shell.  
Note that the predicted visibilities become plateau-like 
at spatial frequencies $\ga 20$~cycles/arcsec 
in Fig.~\ref{risebestfit}b.  This plateau results from the 
unresolved central star.  
It should also be noted that the good match in the wavelength range 
shorter than 1 \mbox{$\mu$m}\ is simply due to the adoption of the empirical 
extinction for a dust cloud, as described in Sect.~\ref{sec_model}.

\begin{figure}
\resizebox{\hsize}{!}{\rotatebox{0}{\includegraphics{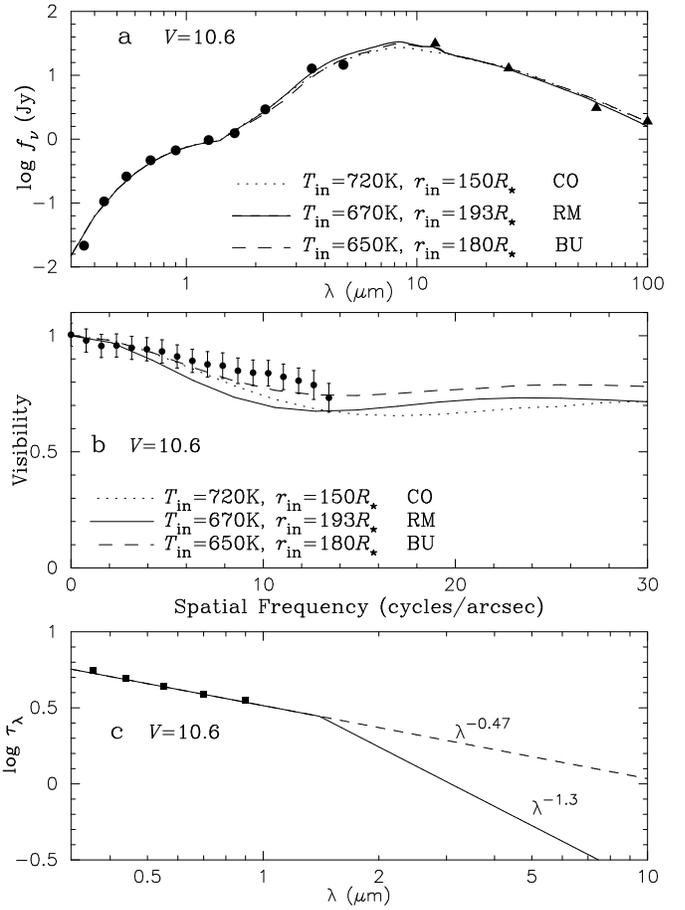}}}
\caption{Simultaneous fit of the SED and visibility of R CrB 
observed on 1999 September 28, 
using an extinction curve different from that extrapolated. 
{\bf a} See the legend to Fig.~\ref{midbestfit}a for the reference of 
the symbols. 
{\bf b} The filled circles represent the observed visibility.  
{\bf c} The solid line represents an extinction curve for the dust 
cloud with a change of the spectral index at 1.4~\mbox{$\mu$m}.  The dashed 
line represents the extinction curve derived by extrapolation 
from the optical depth shortward of 1~\mbox{$\mu$m}, as discussed in 
Sect.~\ref{sec_model}.  The filled squares represent the optical depths 
derived from the observations as shown in Fig.~\ref{extfig}}
\label{sedvisminext}
\end{figure}

Figs.~\ref{midbestfit}a and ~\ref{midbestfit}b reveal 
that the visibility observed at minimum
light cannot be reproduced simultaneously with the observed SED.  
The observed SED can be well fitted using the models with \mbox{$T_{\rm in}$}\ = 
700 -- 800~K and \mbox{$r_{\rm in}$}\ = 116 -- 172~\mbox{$R_{\star}$}, but the visibilities 
predicted from these models are too low as compared with the observation.  
Regarding this discrepancy, 
we first examine the approximations adopted in our models.  
The line-blanketing effect in the atmosphere of RCB stars is very 
prominent as Asplund et al. (\cite{asplund97}) show, but the use of 
line-blanketed atmospheres instead of the blackbody would 
lower \mbox{$T_{\rm in}$}\ by only $\sim 50$~K.  The effect of the uncertainty of the 
effective temperature is also minor.  A decrease of the effective temperature 
by 500~K leads to a decrease of the dust temperature 
by $\sim 60$~K at $\sim 100$~\mbox{$R_{\star}$}.  Combining these effects, 
the dust temperature in the shell can be by $\sim 100$~K lower 
than those used in the fitting above.  We tried to fit the observed 
SED and visibility using such a temperature distribution, but it has 
turned out that the match to the observations is not much improved.

One concern is the estimation of the $K$-band flux of the central star 
obscured by the dust cloud.  As described in Sect.~\ref{sec_model}, 
the extinction of the dust cloud in the $K$-band is derived by 
extrapolation from the region shortward of 1~\mbox{$\mu$m}.  However, 
the actual extinction curve of the dust cloud could become 
steeper longward of 1~\mbox{$\mu$m}\ and approach the usual extinction curve 
of amorphous carbon, which is characterized by a spectral index of 
$\sim 1.3$ (Le Bertre \cite{lebertre97}).  
We find that the observed SED and visibility can be 
simultaneously fitted by adopting an extinction curve for the 
obscuring dust cloud such as shown in Fig.~\ref{sedvisminext}c.  
The extinction curve bends at 1.4~\mbox{$\mu$m}\ and the spectral index 
changes from 0.47 to 1.3.  
Figs.~\ref{sedvisminext}a and ~\ref{sedvisminext}b reveal 
that the observed SED and visibility are fairly reproduced.  
It should be stressed that the physical understanding of such an extinction 
curve is still unclear, and that the exact shape of the extinction 
curve cannot be uniquely determined by the fitting presented here.  
However, the unusual properties of grains in the newly formed dust cloud 
may be responsible for the discrepancy found for the minimum light data.

\subsection{Thermal emission from a newly formed dust cloud}
\label{cloudmodel}

\begin{figure}
\resizebox{\hsize}{!}{\rotatebox{0}{\includegraphics{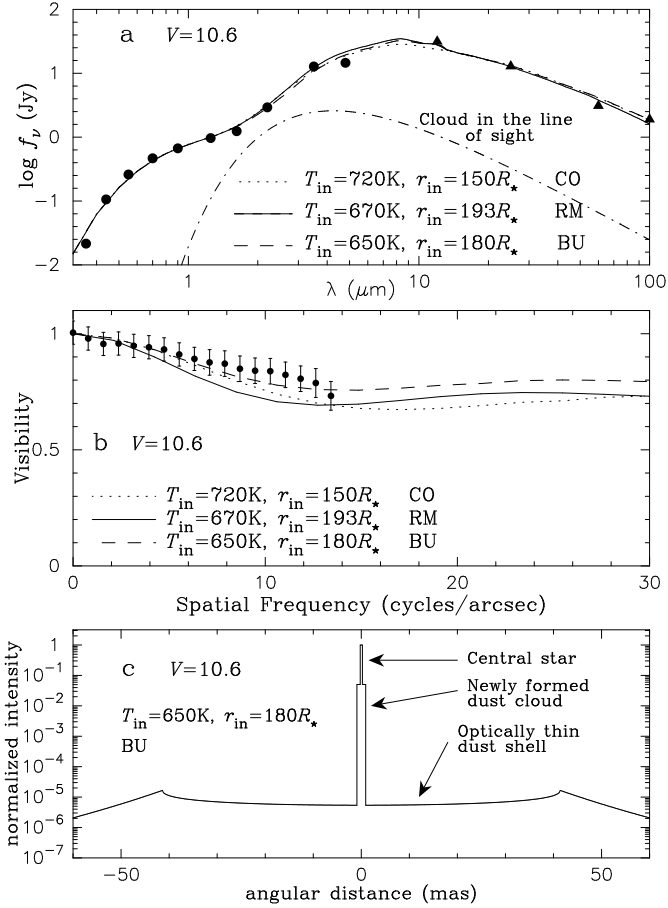}}}
\caption{Simultaneous fit of the SED and visibility of R~CrB 
observed on 1999 September 28, using models with thermal emission from 
a newly formed dust cloud, as discussed in Sect.~\ref{cloudmodel}.  
{\bf a} Three models with different data sets for the opacity of 
amorphous carbon are plotted.  The radius and the temperature of a 
newly formed dust cloud are 4.5~\mbox{$R_{\star}$}\ and 1200~K, respectively.  
CO, RM, and BU denote the opacities of amorphous carbon obtained 
by Colangeli et al. (\cite{colangeli95}), Rouleau \& Martin 
(\cite{rouleau91}), and Bussoletti et al. (\cite{bussoletti87}), 
respectively.  The contribution of the newly formed dust cloud 
is shown with the dash-dotted line for the BU model.  
See the legend to Fig.~\ref{midbestfit}a for the reference of 
the symbols. 
{\bf b} The filled circles represent the observed visibility. 
{\bf c} Normalized intensity profile ($\lambda = 2.1~\mbox{$\mu$m}$) of the 
BU model.  The model consists of the 
central star, a newly formed optically thick dust cloud with 
\mbox{$R_{\rm cl}$}\ = 4.5~\mbox{$R_{\star}$}\ (1.0~mas) and 
\mbox{$T_{\rm cl}$}\ = 1200~K, and an optically thin dust shell with 
\mbox{$r_{\rm in}$}\ = 180~\mbox{$R_{\star}$}\ (41~mas) and \mbox{$T_{\rm in}$}\ = 650~K}
\label{midcloud}
\end{figure}

\begin{figure}
\resizebox{\hsize}{!}{\rotatebox{0}{\includegraphics{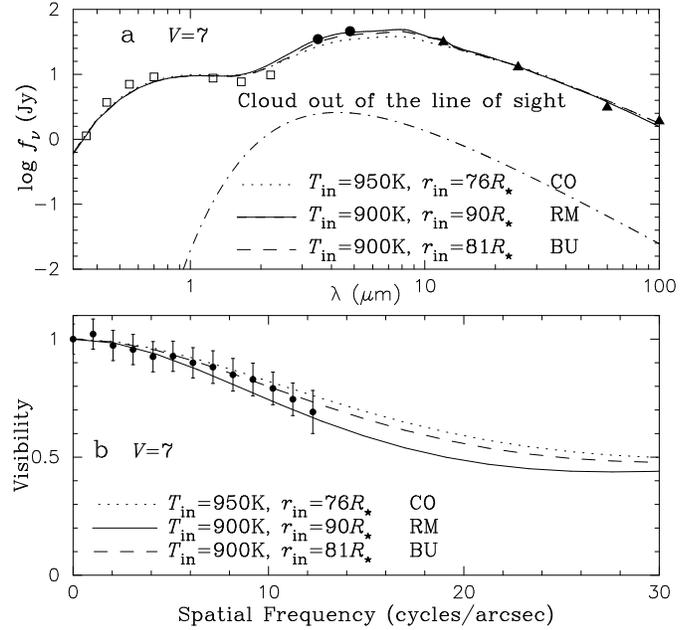}}}
\caption{Simultaneous fit of the SED and visibility of R~CrB 
observed on 1996 October 1, using models with thermal emission from 
a newly formed dust cloud {\em out of} the line of sight, 
as discussed in Sect.~\ref{cloudmodel}.  
A cloud with \mbox{$R_{\rm cl}$}\ = 4.5~\mbox{$R_{\star}$}\ and \mbox{$T_{\rm cl}$}\ = 1200~K is placed out 
of the line of sight, 20~\mbox{$R_{\star}$}\ (5~mas) offset from the central star.  
{\bf a} The three curves represent the SEDs predicted from the models. 
CO, RM, and BU denote the opacities of amorphous carbon obtained 
by Colangeli et al. (\cite{colangeli95}), Rouleau \& Martin 
(\cite{rouleau91}), and Bussoletti et al. (\cite{bussoletti87}), 
respectively. 
The contribution of the newly formed dust cloud 
is shown with the dash-dotted line for the BU model.  
See the legend to Fig.~\ref{risebestfit}a for the reference of 
the symbols. 
{\bf b} The filled circles represent the observed visibility}
\label{risecloud}
\end{figure}

We show an alternative to interpret the data obtained at minimum light.  
We propose that the inconsistency found for the 1999 data 
set may be attributed to the thermal emission 
of a newly formed optically thick dust cloud in front of the star.  
The newly formed dust cloud is supposedly still rather close to the star and 
its angular size is not yet large enough to be resolved with the 6~m
telescope.  
This assumption 
may be reasonable, because the 1999 speckle observation was carried out 
in a decline when the star was obscured by $\Delta V = 4.6$.  
The size and the temperature of such a dust cloud are by no means 
obvious, but the temperature (\mbox{$T_{\rm cl}$}) 
should not exceed $\sim 1500$~K, 
otherwise the thermal emission of the cloud would be too prominent and would 
lead to a poorer match to the observed SED.  
We assume a 
circular cloud (projected onto the sky as seen from the star) 
with a radius of \mbox{$R_{\rm cl}$}\ between 1~\mbox{$R_{\star}$}\ and 10~\mbox{$R_{\star}$}.  
As shown in Fig.~\ref{inset}, \mbox{$R_{\rm cl}$}\  denotes 
the radius of an individual cloud itself, 
not the distance between the cloud and the central star.  
As 1~\mbox{$R_{\star}$}\ corresponds to 0.23~mas, even a cloud with a radius of 
10~\mbox{$R_{\star}$}\ can well be treated as a point source for our speckle 
interferometry.  The thermal emission of the cloud is 
approximated by the blackbody radiation, 
as given by equation (\ref{flcloud}).  
Fig.~\ref{midcloud}c shows the normalized intensity profile of 
a model consisting of the central star, 
an optically thin dust shell, and a thermally emitting optically thick 
dust cloud, whose parameters will be derived below.

Figs.~\ref{midcloud}a and ~\ref{midcloud}b show 
a simultaneous fit using models with 
thermal emission from a newly formed dust cloud.  
The observed SED and visibility are now fairly reproduced with 
models consisting of the central star, an optically thin dust shell 
with an inner radius of \mbox{$r_{\rm in}$}\ = 150 -- 193~\mbox{$R_{\star}$}\ (average
172~\mbox{$R_{\star}$}) and \mbox{$T_{\rm in}$}\ = 650 -- 720~K (average 685~K), 
and an optically thick dust 
cloud with \mbox{$R_{\rm cl}$}\ = 4.5~\mbox{$R_{\star}$}\ and \mbox{$T_{\rm cl}$}\ = 1200~K.  
The fit should be regarded as tentative.  
The uncertainties of \mbox{$T_{\rm in}$}\ and \mbox{$r_{\rm in}$}\ are 
estimated to be $\pm 100$~K and $\pm 20$~\mbox{$R_{\star}$}, respectively, 
while those of \mbox{$T_{\rm cl}$}\ and \mbox{$R_{\rm cl}$}\ are about $\pm 100$~K and 
$\pm 1$~\mbox{$R_{\star}$}, respectively. 

The difference of \mbox{$r_{\rm in}$}\ derived for the 
1996 and 1999 data may indicate a variation of the dust ejection 
frequency.  If the star experiences dust ejection less frequently, 
the inner boundary is expected to become 
larger, and the IR excess is expected to decrease.  In fact, 
as mentioned in Sect.~\ref{sec_photo}, 
the $L$ and $M$ magnitudes decreased monotonically by \mbox{$\sim 1$~mag}  
from 1996 to 1999.  This is consistent with the trend found for 
\mbox{$r_{\rm in}$}\ derived for the two epochs, but further observations are 
required to confirm this correlation.  

Now we apply the thermally emitting cloud models to the data obtained near 
maximum light as well for the following reason. 
Fig.~\ref{lightcurve} shows that R CrB started its final recovery from 
the 1995-96 deep minimum about 6 months before our speckle observation 
on 1996 October 1.  
If dust ejection occurs rather frequently, for example, 
every pulsational cycle (40 -- 50~days), we can expect that the 
star underwent dust ejection \emph{out of} the line of 
sight during these 6 months.  
It is likely that newly formed dust clouds \emph{out of} the line 
of sight existed on 1996 October 1, in addition to the dust cloud 
in front of the star responsible for an obscuration of $\Delta V \sim 1$. 
The latter cloud may already be regarded as part of the optically thin 
dust shell, because $\tau^{\rm cl}_{K} = 0.34$ is derived for this cloud by 
the extrapolation described in Sect.~\ref{sec_model}.  
Therefore, for near maximum light, we consider a model with 
the central star dimmed by $\Delta V \sim 1$, 
thermal emission from an optically thin dust 
shell, and that from a newly formed optically thick dust 
cloud \emph{out of} the line of sight.  
Fig.~\ref{risecloud} shows the SED and 
visibility calculated with an optically thick dust cloud with a radius of 
\mbox{$R_{\rm cl}$}\ = 4.5~\mbox{$R_{\star}$}\ and \mbox{$T_{\rm cl}$}\ = 1200~K, 
in addition to the dimmed central star and the 
optically thin dust shell, whose parameters are the same as shown in 
Fig.~\ref{risebestfit}.  
The cloud is placed out of the line of sight, 20~\mbox{$R_{\star}$}\ (5~mas) 
offset from the central star in the plane of the sky. 
Since the central star has already almost regained its 
brightness near maximum light, the effect of the newly formed dust cloud 
is minor on the SED and visibility.  
Therefore, the models including thermal emission from a newly formed 
dust cloud can also 
provide good matches to the SED and visibility observed 
near maximum light.

\section{Concluding Remarks}

Our 75~mas resolution speckle interferometric observations 
with the SAO 6~m telescope have spatially resolved 
the dust shell around R CrB for the first time.  
Neither the observation near maximum light nor at minimum light 
shows any clear deviation from circular symmetry.  

In order to derive the size of the dust shell, 
we first considered models consisting of the central star 
and an optically thin dust shell, neglecting the thermal emission 
of a newly formed dust cloud.  
Simultaneous fits of the models to the observed SED and visibility 
have demonstrated that a model with the central star 
and an optically thin dust shell with density proportional to 
$r^{-2}$ seems to be appropriate for R CrB near maximum light.  
The inner boundary is found to be $82~\mbox{$R_{\star}$}\ \mbox{(19~mas)} 
\pm 23$~\mbox{$R_{\star}$}\ 
with a temperature of $920 \pm 103$~K.  

This simple picture fails to simultaneously 
reproduce the SED and visibility observed at minimum light, 
which has led us to investigate 
models with thermal emission from 
a newly formed optically thick dust cloud 
whose angular size is not yet large 
enough to be spatially resolvable.  The SED and visibility obtained at 
minimum light were shown to be well fitted with such models.  
The presence of a newly formed dust cloud as hot as 1200~K with 
a radius of 4 -- 5~\mbox{$R_{\star}$}\ is inferred, 
together with an optically thin dust shell 
with $\mbox{$r_{\rm in}$} \sim 170$~\mbox{$R_{\star}$}\ and $\mbox{$T_{\rm in}$}\ \sim 690$~K.  
Furthermore, the SED and visibility obtained near maximum light 
were shown to be fitted also using 
a model with a newly formed dust cloud {\em out of } the line of sight.  
However, we have also discussed that the discrepancy found for the 
minimum light data may be attributed to the unusual 
extinction curve of the obscuring dust cloud.  
Observations during the very bottom of a deep minimum, when the 
contribution of the central star is truly negligible, are 
crucial for investigating the dust shell and dust clouds.

\begin{acknowledgement}
We have used, and acknowledge with 
thanks, data from the AAVSO International Database, 
based on observations submitted to the AAVSO by variable 
star observers worldwide. 
N.R.I. acknowledges the support of the Long-Term Cooperation program of 
the Alexander von Humboldt Foundation.

\end{acknowledgement}

\end{document}